\documentstyle[preprint,aps]{revtex}
\tighten
\draft
\begin{document}
\title{Quantum Measurement in Electric Circuit}
\author{L.S.Levitov$^a$ and G.B.Lesovik$^b$}
\address{(a) Massachusetts Institute of Technology,
12-112, Department of Physics,\\
77 Massachusetts Ave., Cambridge, MA 02139}
\address{(b) Universit\"at zu K\"oln, Institut f\"ur Theoretische Physik,\\
Z\"ulpicher Str. 77, D-5000 K\"oln 41, Germany}
\maketitle

\begin{abstract}
We study fluctuations of electric current in a quantum  resistor
and   derive   a  general  quantum-mechanical  formula  for  the
distribution of transmitted charge.  For  that  we  introduce  a
scheme of current measurement that involves a spin $1/2$ coupled
to  the current so that it precesses at the rate proportional to
the current.  Our  approach  allows  the study of charge  transfer
without  breaking  the  circuit.  We  analyze  a  single channel
conductor and derive electron counting statistics for  arbitrary
relation  between  temperature  and  voltage.  For  a  perfectly
transmitting channel the counting distribution is gaussian, both
for  zero-point  fluctuations  and  at  finite  temperature.  At
constant   voltage  and  low  temperature  the  distribution  is
binomial, i.e., it arises from Bernoulli statistics.

\end{abstract}
\narrowtext

\noindent{\it Introduction.}\hskip4mm
Instantaneous measurement is described in quantum  mechanics  by
wavepacket  reduction,  i.e., by projection on eigenstates of an
observable. A different kind of  measurement,  extended  in  the
time domain, is realized in detectors and other counting
devices. It is known that in such cases a certain revision of
the measurement description is necessary. A famous example from
quantum optics is the theory of photon
detectors\cite{MandelGlauber}. Due to Bose statistics, the
photons entering a photo-counter are correlated in time, and
this makes the theory of photon detection a problem of
many-particle statistics. For a single normal mode of radiation
field the probability $P_m$ to count $m$ photons over time $t$
is given by
   \begin{equation}\label{1}
P_m={(\eta t)^m \over m!}\langle\ :(a^+a)^m e^{-\eta t a^+a}:\ \rangle \ ,
\end{equation}
    where $a^+$ and $a$ are Bose operators of the mode, $\eta$
is "efficiency" of the counter, $\langle ...\rangle$ stands for
the average over a quantum state. The normal ordering $:\dots:$
is important --- physically, it means that, after having been
detected, each photon is destroyed, e.g., it is absorbed in the
detector. Instead of the probabilities, it is more convenient to
deal with the generating function
  \begin{equation}\label{2}\chi(\lambda)=\sum_m \ P_m e^{i\lambda m}\ .
\end{equation}
  For the single normal mode Eq.(\ref{1}) leads to
  \begin{equation}\label{3}\chi(\lambda)=
\langle\ :\exp \eta t (e^{i\lambda}-1) a^+a :\ \rangle \ .
\end{equation}
Eqs.(\ref{1},\ref{2},\ref{3}) account very well for numerous
experimental situations
\cite{Gradinger}. Particularly interesting is the case of a
coherent state $|{\rm z}\rangle$, $a|{\rm z}\rangle=z\ |{\rm z}
\rangle$, where $z$ is a complex number. It corresponds to the
radiation field of an ideal laser, and with Eq.(\ref{3}) one
easily gets Poisson counting distribution,
  \[ P_m={(Jt)^m \over m!}e^{-Jt},\qquad
J=\eta|z|^2\ ,\]
  which describes a minimally bunched light source.

In this paper we discuss measurement of electric current in the
quantum regime, having as a primary goal a complete description
of charge fluctuations, rather than developing measurement
theory for that case. We shall derive a microscopic formula  for
electron counting distribution (see Eqs.(\ref{9}),(\ref{20}))
that can be used for
any  system,  e.g., with an interaction or with a time-dependent
potential. As an example, we find statistics in a single channel
ideal conductor for non-equilibrium  and  equilibrium  noise  at
finite temperature, and for zero-point equilibrium fluctuations.

\noindent{\it Motivation.}\hskip4mm
Similar to the photon detection, electric
measurement is performed on a system containing an enormous
number of particles --- in this case fermions --- and thus one
expects the effects of Fermi statistics to be important. Also,
the duration of electric measurement is typically much longer
than the time it takes the system to transmit one electron by
microscopic tunneling, scattering or diffusion. Apart from these
similarities, there is, however, a crucial difference from the
photon counting: at the detection of a current fluctuation the
number of electrons is not changed, since electric charge is
conserved. This has to be contrasted with absorption of photons
in photo-counters. Related to this, there is another important
difference: at every detection of a photon, its energy
$\hbar\omega$ is taken from the radiation field, which makes
plain photodetectors insensitive to zero-point fluctuations of
electromagnetic field. On the contrary, the
measurement of current fluctuation is usually performed without
changing energy of the system, which makes the zero-point noise
an unavoidable component of any electric
measurement\cite{0-noise}. (Let us emphasize that the difference
has  nothing  to  do  with the type of quantum statistics, Fermi or Bose,
rather it is the difference between the two  kinds  of  measurement,
e.g.,  see  \cite{e-count},  where the counting of fermions with a
counter of "optical" kind was discussed.)

In the classical picture, the measurement gives the charge
$Q(t)=\int_0^t j(t')dt'$ transmitted during the measurement time
$t$. The probabilities $P_m$ of counting $m$ electrons can then
be obtained by averaging $\delta(Q(t)-me)$ over the state of the
system. In a quantum problem electric current is an operator,
and since currents at different moments do not commute, the
operator of transmitted charge $\hat Q(t)=\int_0^t \hat
j(t')dt'$ generally does not make any sense. Instead, since we
are interested in higher-order statistics of current
fluctuations, beyond $\langle \hat j(t)\rangle$ and $\langle [
\hat j(t_1) \hat j(t_2)]_+ \rangle$, in order to compute
electron counting ditribution, we have to include the measuring
system in the quantum Hamiltonian. Our approach is motivated by
the example of the quantum mechanical systems with strong
coupling to macroscopic environment, introduced by Leggett, that
can be treated consistently only by adding the "measuring
environment" to the quantum problem\cite{Leggett}.

For that we introduce a model quantum galvanometer, a spin $1/2$
that precesses in the magnetic field $B$ of the current. For a
classical system, the rate of precession is proportional to
$B(t)$, and $B(t)$ is proportional to the current $I(t)$:
$B(t)=const{\ }I(t)$. Therefore, the precession angle of the
spin directly measures transmitted charge
${\delta}Q={\int_0^t}I(t')dt'$. We adopt the same measurement
procedure for the quantum circuit, i.e., we include in the electron
Hamiltonian the vector potential due to the spin:
   \begin{equation}\label{4}
\vec A(r)= -\mu\ \vec{\hat\sigma}\times\vec\nabla{1\over|r|}\ .
\end{equation}
   Thus we obtain a Hamiltonian describing motion of electrons,
the measuring spin, and their coupling. Now, according to what
has been said, we have to solve dynamics of the spin in the
presence of the fluctuating current, find the distribution of
precession angles, and then interpret it as a distribution of
transmitted charge. Of course, a question remains about the back
effect of the spin on the system, as in any other problem of
quantum measurement. However, as we find below in (\ref{14})
and (\ref{15}),
only the phase of an electron state is affected by the presence
of the spin, not the amplitude. Moreover, the phase will change
only for the transmitted, but not for the reflected wave. As a
result, the probabilities we obtain do not depend on the
coupling constant of the spin. This justifies the assumption
that the spin measures charge transfer in a non-invasive way.

It is worth remarking that our scheme resembles the "Larmor clock"
approach to the problem of traversal time for motion through a
classically forbidden region\cite{L-clock}. In this problem one
is interested, e.g., in the time spent by a particle
tunneling through a barrier. The Larmor clock approach involves
an auxiliary constant magnetic field $B$ added in the
classically forbidden region, and a spin $1/2$ carried by the
particle that interacts with the field: ${\cal
H}_{int}=-\hat\sigma_zB$. The precesson angle of the spin
measures traversal time. In this approach the field $B$ plays a
role somewhat similar to the parameter $\lambda$ in the
generating function (\ref{2}). The main difference from our scheme
is that, in the many-particle problem, instead of letting each
particle carry a spin, we use a single static spin coupled to
current.

\noindent{\it Measurement scheme.}\hskip4mm
Having clarified our motivation, we proceed
semiphenomenologically and choose a new vector
potential in the spin-current interaction $-{1\over c}\vec j
\vec A$. We replace the Amp\`ere's long-range form (\ref{4}) by
a model vector potential
    \begin{equation}\label{5}
\hat A_i(r)= {\lambda\Phi_0\over 4\pi}\ \hat\sigma_z\
\nabla_i\theta(f(r)-f_0)
\end{equation}
   concentrated on some surface $S$ defined by the equation
$f(r)=f_0$. Here $\Phi_0=hc/e$, $\lambda$ is a coupling
constant, $f(r)$ is an arbitrary function, and, as usual,
$\theta(x)=1$ for $x>0$, $0$ for $x<0$.
The surface $S$ defines a section of the conductor on
which the interaction is localized:
    \[
{\cal H}_{int}=
\int -{1\over c}\ \vec{\hat j}\vec{\hat A}\ d^3r=
-{\lambda \hbar\over 2e}\hat\sigma_z\hat I_S\ ,
\]
   where $\hat I_S=
\int_S {\vec {\hat j}}{\vec ds}$, i.e., the spin now is  coupled
to  the  total  current through the section $S$. With the choice
(\ref{5}) of the vector potential one can study  current  fluctuations
in  an  arbitrary section of the conductor. Another advantage of
the phenomenological Eq.(\ref{5}) is that it involves only  one  Pauli
matrix,  which  makes the spin dynamics essentially trivial. The
choice of the quantization axis of the spin is  arbitrary  since
(\ref{5})  will  be  the  only spin-dependent part of the Hamiltonian.
Finally, by switching  from  the  smooth  function  (\ref{4})  to  the
singular  form  (\ref{5}) we enforce integer values of counted charge.
To understand this, let us note that in the "fuzzy" case (\ref{4}) the
measurement can start at the moment when one of the electrons is
located somewhere in the middle of the volume where $A\ne0$, and
then a fractional part of electron charge will  be  counted.  On
the  contrary, in the "sharp" case (\ref{5}), at any time, each electron
can be either on one side of the surface $S$  or  on  the  other
side,  and  thus the  charge  is  detected only when electron's path
crosses $S$. We shall see below in  a  microscopic  calculation,
that  integer  values  of charge follow automatically from gauge
invariance, since the form (\ref{5}) is a gradient of a scalar.

Thus we come to the Hamiltonian
  \begin{equation}\label{7}
\hat{\cal H}_\sigma=\hat{\cal H}(\tilde p, r),\qquad
\tilde p_i= p_i - {e\over c}\hat A_i\ ,
\end{equation}
   where the spin-dependent $\vec{\hat A}$ is taken in the  form
(\ref{5}).  An  essential feature of our approach is that we allow the
constant $\lambda$ of coupling between the spin and the  current
to  vary,  i.e.,  we consider the spin precession as function of
the parameter $\lambda$. The reason is that, unlike  the  photon
counting  problem, our measurement scheme directly generates the
function $\chi(\lambda)$, and then  the  counting  probabilities
$P_m$ are obtained by reading Eq.(\ref{2}) backwards.

At this point we are able to formulate our main result.  Let  us
construct a new
Hamiltonian
  \begin{equation}\label{8}
\hat{\cal H}_\lambda = \hat{\cal H}(\tilde p, r),\qquad
\tilde p_i= p_i - {1\over 2}\lambda\hbar\ \nabla_i\theta(f(r)-f_0)\ ,
\end{equation}
    simply by supressing $\hat\sigma_z$ in Eq.(\ref{5}). The Hamiltonian
$\hat{\cal H}_\lambda$ involves only quantities of the electron
subsystem. Below we show that by studying precession of the spin
coupled to the current, one can get the quantity
  \begin{equation}\label{9}
\chi(\lambda)=\langle e^{i\hat{\cal H}_{-\lambda}t}
e^{-i\hat{\cal H}_{\lambda}t}\ \rangle\ .
\end{equation}
    Here  the  brackets  $\langle...\rangle$ stand for averaging
over initial state of electrons. Note  that  $\chi(\lambda)$  is
written  in  terms  of  a purely electron problem, not involving
spin variables. We shall find that the function  $\chi(\lambda)$
defines the result  of  any measurement of the spin polarization at
the time $t$ when the spin-current  coupling  is  switched off.
Moreover,  we  shall  see that the function (\ref{9}) has the meaning of
a generating function of electron counting distribution, i.e.,
Fourier   transform   of    $\chi(\lambda)$    gives    counting
probabilities, entirely analogous to (\ref{2}).

\noindent{\it Spin dynamics.}\hskip4mm
Our  goal  now will be to express evolution of the spin in terms
of quantities corresponding to the electron system.
The interaction is given by Eqs.(\ref{5}),(\ref{7}).
Suppose that the measurement starts at the moment $0$ and lasts until
time $t$, i.e., the spin-current interaction is switched on
during the time interval $0<\tau<t$. Let us evalute the density
matrix ${\hat {\rho}}_s(t)$ of the spin, right after it is
disconnected from the circuit. We have
  \[{\hat  {\rho}}_s(t)=tr_e(e^{-i{\hat{\cal  H}}_\sigma t} {\hat
{\rho}}e^{i{\hat{\cal H}}_\sigma t})\ ,\]
    where ${\hat {\rho}}$ is initial density matrix ${\hat
{\rho}}_e{\otimes}{\hat {\rho}}_s$ at $t=0$, ${\hat {\rho}}_e$
is initial density matrix of electrons,  and $tr_e(...)$ means
partial trace taken over electron indices, the spin indices left
free. In terms of the spin variables, the operator
$e^{-i{\hat{\cal H}}_\sigma t}$ is a function only of
${\hat\sigma}_z$, and hence it is diagonal in spin:
  ${\langle}{\uparrow}|e^{-i{\hat{\cal H}}_\sigma t}|{\downarrow}{\rangle}=
{\langle}{\downarrow}|e^{-i{\hat{\cal H}}_\sigma t}|{\uparrow}{\rangle}=0$.
  In other words, if initially the spin is in a pure state, up
or down, it will not precess. For ${\hat {\rho}}_s(t)$ this
remark yields:
  \begin{equation}
\label{10}
{\hat {\rho}}_s(t)=\left[\matrix{
  {\hat {\rho}}_{\uparrow \uparrow}(0) &
  {\chi}({\lambda}){\hat {\rho}}_{\uparrow \downarrow}(0) \cr
  {\chi}(-{\lambda}){\hat {\rho}}_{\downarrow \uparrow}(0) &
  {\hat {\rho}}_{\downarrow \downarrow}(0)\cr }\right] .
  \end{equation}
Here $\chi(\lambda)=tr( e^{i\hat{\cal H}_{-\lambda}t} {\hat {\rho}_e}
e^{-i\hat{\cal H}_{\lambda}t} )$, where $ e^{-i\hat{\cal
H}_{\lambda}t} $ is the evolution operator for the problem (\ref{8}). Now,
after the spin degrees of freedom are taken care of by (\ref{10}), we
are left with a purely electron problem, that involves only
electron degrees of freedom but not the spin. By using cyclic
property of the trace one can show that ${\chi}({\lambda})$ in
Eq.(\ref{10}) is identical to (\ref{9}).

In principle, any entry of a density matrix can be measured,
hence the quantity $\chi(\lambda)$ is also measurable. In order
to make clear the relation of $\chi(\lambda)$ with the
distribution of precession angles, let us recall the
transformation rule for the spin $1/2$ density matrix under
rotation by an angle $\theta$ around the $z-$axis:
  \begin{equation}
\label{11}
{\cal R}_\theta(\hat\rho)=\left[\matrix{
  {\hat {\rho}}_{\uparrow \uparrow} &
  e^{-i\theta}{\hat {\rho}}_{\uparrow \downarrow} \cr
  e^{i\theta}{\hat {\rho}}_{\downarrow \uparrow} &
  {\hat {\rho}}_{\downarrow \downarrow}\cr }\right] .
  \end{equation}
By combining this with Eq.(\ref{2})  we write $\hat\rho_s(t)$ as
   \begin{equation}
\label{12}
{\hat\rho}_s(t)=\sum_m P_m{\cal R}_{\theta=m\lambda}(\hat\rho)\ ,
\end{equation}
    which assigns to $P_m$ the meaning of the probability to observe
precession angle $m\lambda$. Let us finally note that such
interpretation of $P_m$ is consistent with what one expects on
classical grounds, because for a {\it classical} magnetic moment
$\vec\sigma$ interacting with the current according to (\ref{5}) the angle
$\theta=\lambda$ corresponds to the precession due to a
current pulse carrying the charge of one electron.

\noindent{\it Single-channel conductor: general formalism.}\hskip4mm
In order to see Eq.(\ref{9}) working,
let us consider an ideal single channel
conductor, i.e., the Schr\"odinger equation
   \begin{equation}\label{13}
i{\partial\psi\over\partial t}= [{1\over
2}(-i{\partial\over\partial x }-{\lambda\over 2}\delta(x))^2+U(x)]\psi
\end{equation}
   in one dimension, where the potential $U(x)$ represents
scattering region and the vector potential is inserted according
to (\ref{5}) at the $x=0$ section. In order to describe
transport, we shall use scattering states, left
and right. Their population $n_{L(R)}(E)$ are equilibrium Fermi
functions with temperature $T$ and chemical potentials shifted
by $eV$, $\mu_L-\mu_R=eV$, thus representing a dc voltage. We
would like to study the case of small $T,eV\ll E_F$, when only
the states close to the Fermi level are important. Such states
have almost linear dispersion law, and thus all wavepackets
travel with the speed $v_F$ without changing shape. In such a
case, instead of the usual scattering states, it is convenient
to assume that the dispersion is stricktly linear and, following
Landauer and Martin\cite{martin}, to use the representation of
time-dependent scattering wave packets
    \[{\psi}_{L,\tau}(x,t)=\cases{
\delta(x_-),
\qquad \qquad
\qquad \qquad
t<\tau \cr
e^{i\lambda/2}A_L\delta(x_-)+B_L\delta(x_+),\ \ t>\tau} ,\]
    \begin{equation}
\label{14}
{\psi}_{R,\tau}(x,t)=\cases{ \delta(x_+),
\qquad \qquad
\qquad \qquad t<\tau \cr
e^{-i\lambda/2}A_R\delta(x_+)+B_R\delta(x_-),\ \ t>\tau} ,\end{equation}
  where $x_\pm=x\pm v_F(t-\tau)$. Here $\tau$ is the moment of
scattering of a packet, $A_{L(R)}$ and $B_{L(R)}$ are scattering
amplitudes at $\lambda=0$. To make expressions less heavy, we
drop electron spin. Also, we assume that the scattering
amplitudes are energy-independent, which is equivalent to
replacing the barrier $U(x)$ by $U_0{\delta}(x)$ and is
consistent with the closeness to $E_F$. The phase factors
$e^{\pm i\lambda/2}$ in (\ref{14}) are found immediately, by observing
that the vector potential in the Schr\"odinger equation can be
eliminated by the gauge transformation $\psi(x) \rightarrow
\exp(i\lambda / 2\ \theta(x)) \psi(x)$. Scattering amplitudes
form a unitary matrix:
  \begin{equation}
\label{15}
\hat S_\lambda =\left[\matrix{
{e^{i\lambda/2}A_L}&
B_R \cr
B_L&
{e^{-i\lambda/2}A_R}\cr }\right]
\end{equation}
    Second-quantized, electron states (\ref{14}) lead to
${\hat {\psi}}(x,t)={\hat {\psi}}_L(x,t)+{\hat {\psi}}_R(x,t)$ with
   \begin{equation}
\label{16}
{\hat {\psi}}_{L(R)}(x,t)={\sum\limits_\tau}{\psi}_{L(R),\tau}(x,t)
{\hat c}_{1(2),\tau}\ ,
\end{equation}
    where $c_{1,\tau}$ and $c_{2,\tau}$ are canonical Fermi
operators corresponding to the states (\ref{14}), the left
and the right respectively. This defines ${\psi}(x)$ in (\ref{13}).

In the basis of the wavepackets (\ref{14}),(\ref{16}) it is straightforward
to write the many-particle evolution operator through the
single-particle scattering matrix ${\hat S}_\lambda$:
   \begin{equation}
\label{17}
e^{-i{\hat {\cal H}}_{\lambda}t}= \exp\int_0^t d\tau \sum\limits_{ij}
\ln[{\hat S}_\lambda]_{ij} c_{i,\tau}^+c_{j,\tau} \ ,
\end{equation}
   To verify (\ref{17}) let us note that in the wavepacket
representation (\ref{14}) Fermi correlations occur only for the pairs
of left and right states that scatter at the same instant of
time. For each of such pairs the evolution operator $e^{-i{\hat
{\cal H}}_{\lambda}t}$ is $\hat 1$ if both states are occupied
or both are empty, otherwise it is given by the single-particle
scattering matrix (\ref{15}).

Using similar arguments, we compute
   \begin{equation}
\label{18}
e^{i{\hat {\cal H}}_{-\lambda}t} e^{-i{\hat {\cal H}}_{\lambda}t}=
\exp\int_0^t d\tau \sum\limits_{ij} {\hat W}_{ij}
c_{i,\tau}^+c_{j,\tau}\ ,
\end{equation}
   where $e^{\hat W}= {\hat S}_{-\lambda}^{-1} {\hat S}_\lambda$
is readily obtained from (\ref{15}):
  \begin{equation}
\label{19}
e^{\hat W}=
\left[\matrix{
{e^{i\lambda}|A_L|^2+|B_L|^2}&
2i\sin\lambda\bar A_LB_R \cr
2i\sin\lambda\bar B_RA_L &
{e^{-i\lambda}|A_R|^2+|B_R|^2} \cr } \right]
\end{equation}
   Using unitarity of $e^{\hat W}$ and commutation rules for
$c_{\alpha,\tau}$ one can rewrite (\ref{18}) in terms of normal
ordering:
   \begin{equation}
\label{20}
e^{i{\hat {\cal H}}_{-\lambda}t} e^{-i{\hat {\cal H}}_{\lambda}t}=\
:\exp\int\limits_0^t d\tau \sum\limits_{ij} [e^{\hat W}-1]_{ij}
c_{i,\tau}^+c_{j,\tau}:\ .
\end{equation}
   This form is ready to be plugged into Eq.(\ref{9}) and averaged
over initial state. Let us note the striking similarity of the two
formulas obtained by different means: the fermionic Eq.(\ref{20}) and the
bosonic Eq.(\ref{3}). Also, let us mention that the periodicity of the
matrix (\ref{19}) in $\lambda$ ensures periodicity of $\chi(\lambda)$,
and thus guarantees integer values of charge.

\noindent{\it Single-channel conductor: results.}\hskip4mm
Let us start with a simple example of a single particle in the
state $c^+_{1,\tau}|{\rm vac}\rangle$ that corresponds to
scattering at the moment $\tau$. In this case, from (\ref{20})
and (\ref{9}) one gets $\chi(\lambda)= e^{i\lambda}|A|^2+|B|^2$
for $0<\tau<t$, $1$ otherwise. ($|A|=|A_L|=|A_R|$,
$|B|=|B_L|=|B_R|$) Evidently, according to Eq.(\ref{2}), this
simply means that for the scattering occurring during operation
of the detector, the counting probabilities are identical to the
one-particle scattering probabilities, as it should be expected.

Now, let us consider current fluctuations in an
equilibrium Fermi gas. Assume perfect
transmission: $B_{L(R)}=0$. Then Eq.(\ref{19}) gives $\hat
W=i\lambda\sigma_z$, and thus Eq.(\ref{18}) becomes
   \begin{equation}
\label{21}
e^{i{\hat {\cal H}}_{-\lambda}t} e^{-i{\hat {\cal H}}_{\lambda}t}=
\exp i\lambda
\int_0^t (c_{1,\tau}^+c_{1,\tau}- c_{2,\tau}^+c_{2,\tau}) d\tau \ ,
\end{equation}
   i.e., the right and the left states separate. We observe that
the averaging of (\ref{21}) over the Fermi ground state is identical
to that performed in the orthogonality catastrophe
calculation\cite{ortho}, so we get
   \begin{equation}
\label{22}
\chi(\lambda)= e^{-\lambda^2f(t,T)}\ ,
\end{equation}
$f(t,T)=\langle\!\langle(\int_0^t
c_{1,\tau}^+c_{1,\tau}d\tau)^2\rangle\!\rangle=
\cases{ {1\over2\pi^2}\ln E_Ft,\ t\ll\hbar/T \cr
Tt/h,\ t\gg\hbar/T }$.
   According to (\ref{2}), this leads to gaussian counting statistics.

Let us remark that, incidentally, Eq.(\ref{22}) also gives a solution to
another problem: the statistics of the number of fermions inside
a segment of fixed length in one dimension. The relation is
immediately obvious after one assigns to $\tau$ in Eq.(\ref{21}), the
meaning of a coordinate on  a line. Thus, the statistics are gaussian.

Now, it turns out that the general case $B\ne0$ can be
reduced to (\ref{21}) by a canonical transformation
of $c_{\alpha,\tau}$ that makes the quadratic form in (\ref{18})
diagonal. The transformation is related in
the usual way with the  eigenvectors
of the matrix $\hat W$. Thus, we come to Eqs.(\ref{21}),(\ref{22}) with
$\lambda$ replaced by $\lambda_*$:
$\sin{\lambda_*\over2}=|A|\sin{\lambda\over2}$. The counting
statistics in this case are non-gaussian:
   \begin{equation}
\label{23}
\chi(\lambda)= e^{-\lambda_*^2f(t,T)}\ .
\end{equation}
   One checks that the second moment of the distribution agrees
with the Johnson-Nyquist formula for the equilibrium noise.

Finally, let us find statistics for the non-equilibrium noise.
In this case, due to the asymmetry in the population,
$n_{L(R)}(E)=(\exp (E{\pm}{1\over2}eV)/T+1)^{-1}$, generally one
cannot uncouple the two channels by a canonical transformation.
We calculate the statistics within an approximation that
neglects by the effect of switching at $\tau=0$ and $\tau=t$.
Let us close the axis $\tau$ into a circle of
length $t$, which amounts to restricting on periodic states:
${\psi}(\tau)={\psi}(\tau\pm t)$. For the $t-$periodic problem,
by going to the Fourier space, one has
  \[{\chi}({\lambda})=
{\prod\limits_k} [1+|A|^2(e^{-i{\lambda}}-1)n_L(E_k)(1-n_R(E_k))+
|A|^2(e^{i{\lambda}}-1)n_R(E_k)(1-n_L(E_k))]\ ,
\]
   where $E_k=2\pi\hbar k/t$, $k$ is an integer. For large $t$,
$t\gg\hbar/T$ or $t\gg\hbar/eV$, the product is converted to an
integral:
  \[\ln({\chi}({\lambda}))=
{t\over 2{\pi}{\hbar}}{\int_{-\infty}^{+\infty}}
dE\ \ln(1+|A|^2(e^{-i{\lambda}}-1)n_L(1-n_R)+
|A|^2(e^{i{\lambda}}-1)n_R(1-n_L))\ .\]
   We evaluate it analytically, and get
\begin{equation}
\label{24}
{\chi}({\lambda})=
\exp -tT u_+u_-/h\ ,
\end{equation}
  where $u_\pm=v\pm\cosh^{-1}(|A|^2\cosh(v+i\lambda)+|B|^2\cosh v)$,
$v=eV/2T$.
The answer
simplifies in the two limits: $T\gg eV$ and
$eV\gg T$.
In the first case we return to the equilibrium result (\ref{23}). In
the second case, corresponding to the recently discussed quantum
shot noise\cite{qsn}, we have
  \begin{equation}
\label{25}
{\chi}({\lambda})=(e^{i\epsilon\lambda}|A|^2+|B|^2)^{e|V|t/h},
\qquad \epsilon=sgn(V)\ ,
\end{equation}
   Analyzed according to Eq.(\ref{2}), this
${\chi}({\lambda})$ leads to the binomial distribution
 \[P_N(m)=p^mq^{N-m}C_N^m\ ,\]
$p=|A|^2$, $q=|B|^2$, $N=e|V|t/h$. One checks that the moments
${\langle}k{\rangle}=pN$ and
$\langle\!\langle k^2\rangle\!\rangle=pqN$ correspond
directly to the Landauer formula and to the formula for the
intensity  of the quantum shot noise\cite{qsn}.
The correction to the statistics due to  the  switching  effects
is insignificant \cite{LL}.

In conclusion, we introduced a quantum-mechanical scheme that
gives complete statistical description of electron transport.
It involves a spin $1/2$ coupled to the current so that the spin
precession measures transmitted charge. The off-diagonal part of
the spin density matrix, as a function of the coupling constant,
gives generating function for the electron counting statistics.
We find the statistics in a single-channel ideal conductor for
arbitrary relation between temperature and voltage. For a
perfectly transmitting channel the counting distribution is
gaussian, both for zero-point fluctuations and at finite
temperature. At constant voltage and low temperature the
distribution is binomial.

\acknowledgements
Research of L.L. is  partly  supported
by Alfred Sloan fellowship.


\begin{references}
\bibitem{MandelGlauber}
R. J. Glauber, Phys. Rev. Lett. {\bf 10}, 84
(1963); Phys. Rev. {\bf 130}, 2529 (1963);\\
L. Mandel, E. Wolf, Rev. Mod. Phys. {\bf 37}, 231 (1965)
\bibitem{Gradinger} C. W. Gardinger, Quantum Noise, Chapter 8,
Springer-Verlag (1991);\\
J. R. Klauder, E. C. G. Sudarshan, Fundamentals of Quantum Optics,
Chapter 8, W.A.Benjamin, Inc., N.Y. (1968)
\bibitem{0-noise} R. H. Koch, D. van Harlingen, J. Clarke,
Phys. Rev. B{\bf 26}, 74 (1982)
\bibitem{e-count} S. Saito, et al., Phys. Lett. A{\bf 162}, 442 (1992)
\bibitem{Leggett} A.J.Leggett, Progr. Theor. Phys. Suppl.{\bf 69}, 80 (1980)
\bibitem{L-clock} M. B\"uttiker, Phys. Rev. B{\bf 27}, 6178 (1983);\\
A. I. Baz', Sov. Phys. JETP, {\bf 20}, 1261 (1965);\\
A. I. Baz', Ya. B. Zeldovich and A.  M.  Perelomov,  Scattering,
Reactions and Decay in Nonrelativistic Quantum Mechanics, Israel
Program for Scientific Translations, Jerusalem (1969)
\bibitem{martin} Th. Martin and R. Landauer, Phys.Rev. B{\bf 45}, 1742 (1992)
\bibitem{ortho}
P. Nozi\`eres, C. T. deDominicis, Phys.Rev., {\bf 178}, 1084 (1969);\\
K. D. Schotte and U. Schotte, Phys.Rev., {\bf 182}, 479 (1969);\\
G. D. Mahan, Many Particle Physics, Sec. 8.3,
(2-nd edition, Plenum Press, 1990)
\bibitem{qsn}
G.B.Lesovik, JETP Letters, {\bf 49}, 594 (1989);\\
B.Yurke and G.P.Kochanski, Phys.Rev.{\bf 41}, 8184 (1989);\\
M. B\"uttiker, Phys.Rev.Lett., {\bf 65}, 2901 (1990);
\bibitem{LL}
L.S.Levitov, G.B.Lesovik, JETP Letters {\bf 58} (3), 230 (1993)
\end{references}
\end{document}